\documentclass[11pt, authoryear]{elsarticle}              
\journal{arXiv.org}            

\usepackage{amssymb,amsfonts,amsmath}
\usepackage{xcolor}
\usepackage{natbib}
\usepackage[top=1.0in, bottom=1.0in, left=1.0in, right=1.0 in]{geometry}   %
\usepackage[onehalfspacing]{setspace}
\usepackage[bottom]{footmisc}
\usepackage{indentfirst}
\usepackage{endnotes}
\usepackage{verbatim}                
\usepackage{lastpage,placeins}
\usepackage{hyperref}                
\usepackage{graphicx,epsfig,epstopdf,subfig}       
\usepackage{array}
\usepackage{rotating,longtable,booktabs, colortbl}       
\usepackage{mdwlist}
\usepackage{enumerate}
\usepackage[shortlabels]{enumitem}
\usepackage{lscape,afterpage}                    
\usepackage{appendix}
\usepackage{breakcites}
\usepackage{bm}

\usepackage[standard]{ntheorem}
\usepackage[algo2e]{algorithm2e}

\allowdisplaybreaks[1]                     
\theoremstyle{break}
\theorembodyfont{\normalfont}
\newtheorem{algorithm}[algocf]{Algorithm}

\newcommand{\ack}[1]{%
  \nonumnote{\textit{Acknowledgements:\enspace}#1}}


\begin{document}

\begin{frontmatter}
\title{\textbf{Do Online Courses Provide an Equal Educational Value
Compared to In-Person Classroom Teaching? Evidence from US Survey Data using Quantile Regression}}


\author[add1]{Manini Ojha}
\ead{mojha@jgu.edu.in}
\author[add2]{Mohammad Arshad Rahman\corref{cor1}}
\ead{marshad@iitk.ac.in}

\cortext[cor1]{Please address correspondence to Mohammad Arshad
Rahman,
Department of Economic Sciences, Indian Institute of Technology, Kanpur.
Room
672, Faculty Building, IIT Kanpur, Kanpur 208016, India. Phone: +91
512-259-7010. Fax: +91 512-259-7500.}
\address[add1]{Jindal School of Government and Public Policy,
O.P. Jindal Global University, Sonipat, India.}
\address[add2]{Department of Economic Sciences, Indian Institute of
Technology Kanpur,
India.}

\ack{}

\begin{abstract}
Education has traditionally been classroom-oriented with a gradual growth of
online courses in recent times. However, the outbreak of the COVID-19
pandemic has dramatically accelerated the shift to online classes.
Associated with this learning format is the question: what do people think
about the educational value of an online course compared to a course taken
in-person in a classroom? This paper addresses the question and presents a
Bayesian quantile analysis of public opinion using a nationally
representative survey data from the United States. Our findings show that
previous participation in online courses and full-time employment status
favor the educational value of online courses. We also find that the older
demographic and females have a greater propensity for online education. In
contrast, highly educated individuals have a lower willingness towards
online education vis-à-vis traditional classes. Besides, covariate effects
show heterogeneity across quantiles which cannot be captured using probit
or logit models.

JEL codes: A20, C11, C31, C35, I20, I29

\end{abstract}

\begin{keyword} Bayesian quantile regression, binary quantile regression,
COVID-19, educational value, Gibbs sampling, public opinion, Pew Research
Center.
\end{keyword}
\end{frontmatter}


\section{Introduction}\label{sec:Intro}

Advancements in technology have resulted in a proliferation of online
educational opportunities over the last two decades.
\citet{Allen-Seaman-2016} report that the growth rate of enrolments in online
courses is expanding faster than the traditional classroom enrolments in the
United States (US).\footnote{Roughly one in two individuals who have
graduated in the last ten years have taken at least one online course in
their degree program \citep{Parker-etal-2011}.}  Even as academic leaders
remain far more positive about traditional and blended pedagogical formats
than fully online learning, the outbreak of the COVID-19 pandemic has acted
as a catalyst in establishing online education as an immediate substitute for
in-person classrooms. This paradigm shift in education has drawn considerable
attention from the media, and researchers across the globe. However, any
research based on the acceptance of online education during this period of
disturbance is likely to be a deviation from the natural relationship between
education and technology. Instead, as institutions of higher learning
integrate web-based tools into classroom instructions, we assert that it is
more important to assess the attitude towards digital education and its
acceptance in a state of equilibrium prior to the onset of the pandemic. With
that in mind, in this paper, we analyse public opinion on the value of online
education relative to traditional education using survey data from the Pew
Social Trends and Demographics Project conducted by the Princeton Survey
Research Associates International in 2011.


In the early 2000s, despite significant skepticism from academics and
pushback from the public, several universities invested in and adopted
Massive Open Online Course (MOOCs) as a teaching-learning format
\citep{Miller-2014}. Educational institutions today are compelled to rethink
their pedagogical philosophies to incorporate either hybrid or fully-online
teaching-learning formats, as a consequence of the ongoing pandemic. Students
graduating in the current era have experienced some education using
technology, either as a supplement to traditional classes, or as fully online
courses. Correspondingly, faculty is expected to have the willingness and the
ability to engage in pedagogy that utilises technology \citep{Miller-2014}.
While this trend towards instructional technology expands, there are
ambivalent perceptions about the quality of online education
\citep{Chen-etal-2013,Otter-etal-2013,Allen-Seaman-2011}.
Therefore, we specifically address public opinion about the value of online
education and the factors that influence it vis-à-vis traditional classes,
using a Bayesian quantile analysis.


Modelling public opinion on the value of online education presents a rich
area for further study. Over the last few years, a sizeable body of
literature on the demand and efficacy of online education, its scope to lower
educational costs, student and faculty perceptions, and its impact on student
learning outcomes have emerged
\citep{Xu-Jaggars-2013,Goodman-etal-2019,Otter-etal-2013,Cassens-2010,Bettinger-etal-2017,Figlio-etal-2013,Alpert-etal-2016,Joyce-etal-2015,Krieg-Henson-2016,Kirtman-2009}.
However, much of the existing research focuses on one or two specific
courses, or are limited within a selective college or university. For
instance, \citet{Goodman-etal-2019}  compare an online and in-person degree
in Master of Science in Computer Science offered at Georgia Tech and document
a large demand for the online program with nearly no overlap in the applicant
pools. Analysing survey data from a community college in California,
\citet{Cassens-2010} finds no significant differences in students
performances in online and traditional teaching methods. Contrary to this,
using data from one large for-profit university, \citet{Bettinger-etal-2017}
find negative effects of online courses on student academic success and
progression relative to in-person courses. On similar lines,
\citet{Otter-etal-2013} find significant differences upon comparison of
faculty and student perceptions of online courses versus traditional courses
at a large public university in the south-eastern United States. As such,
mixed evidence found owing to the narrow focus of these papers often brings
their external validity into question. To this end, we attempt to address the
educational value of online classes by utilising a nationally representative
US survey data, thereby drawing conclusions for a population at large. This
is the first contribution of our paper to the existing literature on online
education.

Evidence finds the proportion of faculty who believe in the legitimacy of
online education to be relatively low. In addition, the proportion of faculty
who perceive online education as more time intensive and requiring greater
effort has seen a steady growth \citep{Allen-Seaman-2011}. Contrary to this,
students perceive such courses to be largely self-taught with minimal effort
from the faculty \citep{Chen-etal-2013,Otter-etal-2013}. Such ambiguous views
on the issue makes it imperative to investigate overall public opinion on the
matter. That being the case, our paper also contributes to a second body of
literature that points towards the differential adoption and acceptance of
technology in higher education across different demographies
\citep{cooper2006digital,Norum-Weagley-2006,Chen-Fu-2009,Cotten-Jelenewicz-2006,Odell-etal-2000,Jones-etal-2009}.
Given that online education is a matter of individual selection, individual
characteristics may vary drastically across the utility derived from it.
Traditional mean regression of the effects of covariates on the preference
about online classes may mask important heterogeneity in individual choices.
Our study is the first, of which we are aware, to offer new insights
regarding the opinion on the educational value of online courses across the
quantiles and latent utility scale. These differential effects across the
latent utility scale may be of direct interest to policy makers and
educationists as our methodology provides a more comprehensive picture.

Utilising a nationally representative US survey data from the Pew Social
Trends and Demographics Project conducted in the year 2011, we examine
individual responses about the \textit{educational value derived from online
classes in comparison to in-person classroom}. Our paper presents an
empirical application of binary quantile regression to a model of educational
decision. More specifically, we model the latent utility differential between
online classes and traditional classes. This may be interpreted as a
\textit{propensity} or a \textit{willingness} index, where higher propensity
towards online education are characterised by large positive values and vice
versa. The results are compelling and ought to serve as a guide for future
research. We find that an older demographic, individuals with full-time
employment, individuals with previous online experience, and females display
a propensity towards online education. Interestingly, our findings highlight
that highly  educated respondents have lower willingness for online
education. We also note some amount of regional differences in the propensity
to value online classes. All these covariates show considerable differences
in covariate effects at different quantiles. Lastly, we find no convincing
evidence of race or income having an effect on the propensity for online
education.

The remainder of the paper is organised as follows. Section~\ref{sec:Data}
outlines the data used for our analysis including a descriptive summary. This
is followed by Section~\ref{sec:Model} that  outlines a model of quantile
regression for binary outcomes and presents a Markov chain Monte Carlo (MCMC)
algorithm for its estimation. Next, we present the results of our binary
quantile regressions in Section~\ref{sec:Results}.
Section~\ref{sec:Conclusion} presents the concluding remarks.

\section{Data}\label{sec:Data}

The study utilises a nationally representative US survey data from the Pew
Social Trends and Demographics Project, conducted over telephone between
March $15-29$, 2011, by the Princeton Survey Research Associates
International. The survey was primarily for higher education and housing and
contains information on 2,142 adults living in the continental US. We
consider a subset of variables from this survey and upon removing missing
observations from our variables of interest (see
Table~\ref{Table:DataSummary}), we are left with 1,591 observations available
for the analysis. The dependent variable is the response to the question:
\textit{``In general, do you think a course taken only online provides an
equal educational value compared with a course taken in person in a
classroom, or not?''}. Responses are recorded either as \textit{``Yes''},
\textit{``No''}, or \textit{``Don't know/Refused''}. We ignore the last
response category. Of the 1,591 respondents, 505 (31.74\%) respondents agree
that a course taken online provides an equal educational value compared to
in-person classroom teaching, while the remaining 1086 respondents (68.26\%)
do not agree and thus believe that online courses have lesser educational
value. The survey also consists of information on an array of other
variables, some of which we utilise as covariates (independent variable) in
our analysis. A description of the covariates and the response variable,
along with the main characteristic of the data is presented in
Table~\ref{Table:DataSummary}.

\begin{table}[!t]
\centering \footnotesize \setlength{\tabcolsep}{6pt} \setlength{\extrarowheight}
{1.5pt}
\setlength\arrayrulewidth{1pt}
\caption{Descriptive summary of the variables.}
\begin{tabular}{lp{8.5cm}r r}
\toprule
\textsc{variable}      & \textsc{description}   & \textsc{mean}   & \textsc{std}  \\
\midrule
Age/100                 & Age (in years) divided by 100   &  0.44   &  0.18       \\
Income/100,000          & Mid-point of income category (in US dollars) divided
by 100,000
                                                          &  0.63   &  0.48       \\
\midrule
                        &     & \textsc{count} & \textsc{percent}  \\
                        \midrule
Online Course           & Indicates that the respondent has previously taken an
                          online course for academic credit
                                                          &    352   &    22.12   \\
(Age$< 65$)$\ast$ Enroll
                        & Indicates that the respondent is of age below 65 and
                          currently enrolled in school
                                                          &    291   &    18.29  \\
Female                  & Indicator variable for female gender
                                                          &    815   &    51.23   \\
Post-Bachelors          & Respondent's highest qualification is Masters,
                          Professional or Doctorate       &    221   &    13.89   \\
Bachelors               & Respondent's highest qualification is Bachelors
                                                          &    356   &    22.38   \\
Below Bachelors         & Respondent holds a 2-year associate degree, went
to
                          some college with no degree, or attended technical,
                          trade or vocational school after high school
                                                          &    482   &    30.30   \\
HS and below            & Respondent is a high school (HS)
                          graduate or below               &    532   &    33.44   \\
Full-time               & Indicator for full-time employment
                                                          &    757   &    47.58   \\
Part-time               & Indicator for part-time employment
                                                          &    240   &    15.08   \\
Unemployed              & Indicator for either unemployed, student or retired
                                                          &    594   &    37.34   \\
White                   & Indicator for a White respondent
                                                          &   1131   &    71.09   \\
African-American
                        & Indicator for an African-American respondent
                                                          &    257   &    16.15   \\
Other Races
                        & Indicator for a respondent who is either an Asian,
                          Asian-American or belongs to some other race
                                                          &    203   &    12.76   \\
Urban                   & Lives in an urban region        &    626   &    39.35   \\
Suburban                & Lives in a suburban region
                                                          &    760   &    47.77   \\
Rural                   & Lives in a rural region
                                                          &    205   &    12.88   \\
Northeast               & Lives in the Northeast
                                                          &    220   &    13.83   \\
West                    & Lives in the West
                                                          &    362   &    22.75   \\
South                   & Lives in the South
                                                          &    724   &    45.51   \\
Midwest                 & Lives in the Midwest
                                                          &    285   &    17.91   \\
\midrule
Opinion                 & Respondent answered `Yes' to our question of interest
                                                          &   505    &   31.74    \\
                        & Respondent answered `No' to our question of interest
                                                          &  1086    &   68.26    \\
\bottomrule
\end{tabular}
\label{Table:DataSummary}
\end{table}

In our sample, a typical individual is 44 years of age with a family income
of 63 thousand US dollars. The survey recorded income as belonging to one of
the following nine income categories: $ < 10k$, $10k-20k$, $20k-30k$,
$30k-40k$, $40k-50k$, $50k-75k$, $75k-100k$, $100k-150k$ and $>150k$, where
$k$ denotes a thousand dollars. We use the mid-point of each income category
to represent the income variable, where \$5,000 and \$1,75,000 are used as
the mid-point for the first and last income categories, respectively. With
respect to online learning, we have a little more than one-fifth of the
sample who have previously taken an online course for academic credit. A
sizeable proportion of the sample, therefore, have had prior exposure to
online learning. Individuals who are aged less than 65 and currently enrolled
in school comprise a little less than one-fifth of the sample. Here,
enrolment in school implies that the respondent is either attending high
school, technical school, trade or vocational school, is a college
undergraduate or in graduate school.

The sample has more females (51.23\%) than males (48.77\%), but both genders
have approximately equal representation. Education has been classified into
four categories with `High School (HS) and below' forming the largest category (33.44\%)
followed by `Below Bachelors' (30.30\%). The smallest two educational
categories are `Bachelors' (22.38\%) and `Post-Bachelors' (13.89\%). So,
approximately two-thirds of the sample have less than bachelors education.
With respect to employment status, about a little less than two-thirds (i.e.,
62.66\%) are either employed full-time or part-time, while the remaining
percentage are either unemployed, students or retired individuals. Racial
classification shows that more than two-thirds are White (71.09\%), followed
by African-Americans (16.15\%) and all other races (12.76\%). In terms of
rural-urban classification, most of the sampled individuals live in the
suburban areas (47.77\%), followed by the urban areas (39.35\%). The lowest
proportion lives in the rural areas (12.88\%). Regional classification as
defined by the US Census Bureau shows that the largest percentage of the
sample live in the South (47.77\%). This is followed by the West (22.75\%),
Midwest (17.91\%), and Northeast (13.83\%) regions.

\begin{figure}[!t]
	\centerline{
		\mbox{\includegraphics[width=6.75in, height=5in, trim = {0 0.5cm 0
0.5cm},
              clip]{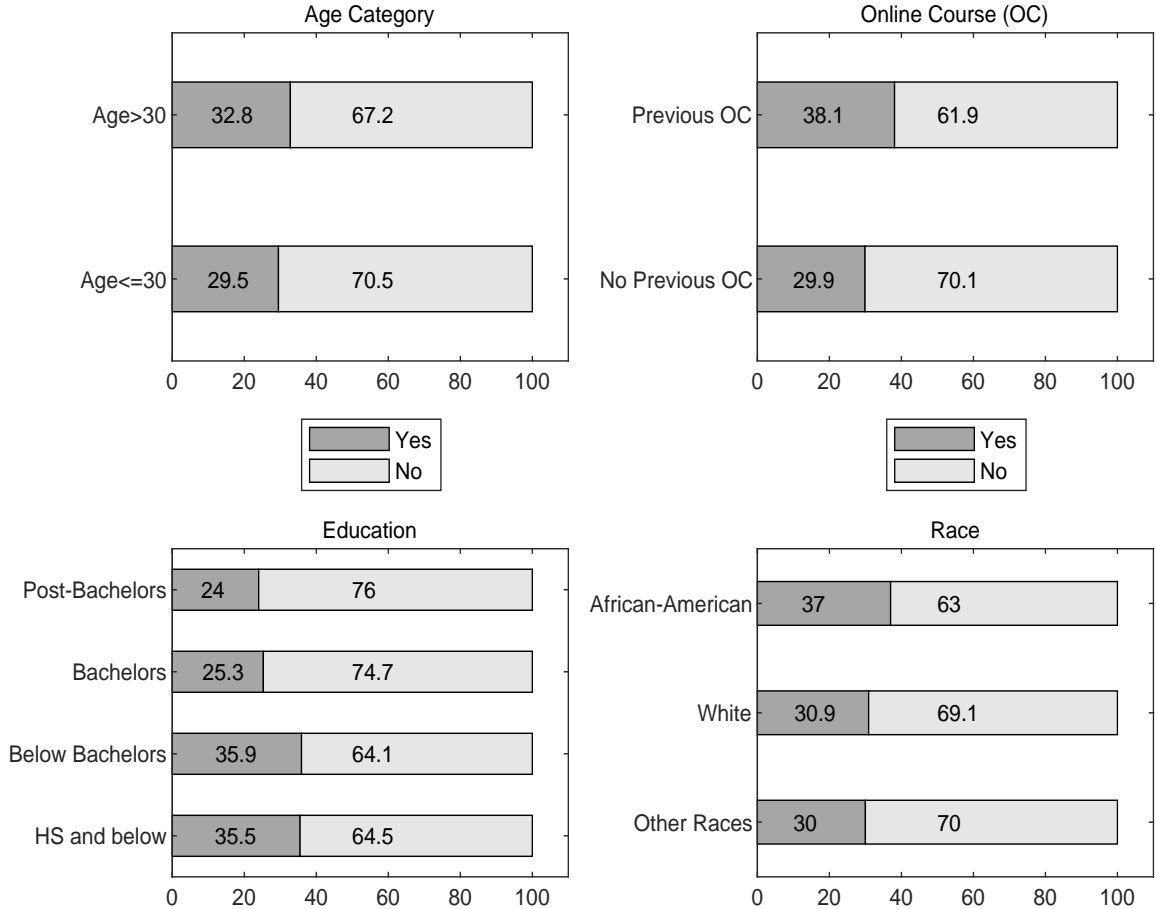}} }  
\caption{Stacked bar graph displaying the percentage of observations
corresponding
to the two categories of public opinion (Yes and No) for each category of
some selected covariates.  }
\label{fig:BarGraph}
\end{figure}

Before we formally delve into modelling the dependent variable (i.e., public
opinion on educational value of online learning relative to in-person
classroom teaching), we explore its relationship with some selected
independent variables or covariates \citep[see][for a report on data
summary]{Parker-etal-2011}. To explore this association, we present a stacked
bar graph in Figure~\ref{fig:BarGraph} with four panels, each portraying the
relationship between the dependent variable and a single covariate. Each bar
within a panel corresponds to a category of the covariate and displays the
percentage of observations that says `Yes' and `No' to our question of
interest. For example, the upper (lower) bar in Panel~1 shows that for people
aged greater than (less than equal to) 30, 32.8\% (29.5\%) of the sample
agree that online courses have the same educational value as in-person
classroom teaching, while the remaining 67.2\% (70.5\%) do not agree. The
other three panels of Figure~\ref{fig:BarGraph} can be interpreted
analogously.

We see from the first panel of Figure~\ref{fig:BarGraph} that the percentage
of sample who says `Yes' (and thus `No') is approximately equal amongst the
younger ($\mathrm{Age} <=30$) and older ($\mathrm{Age} >30$) population. From
Panel~2 we note that, amongst the sample who have taken an online course for
academic credit, a higher percentage (38.1\%) says `Yes' compared to those
(at 29.9\%) who have no previous online learning experience. Panel~3 suggests
that the highly educated group (Bachelors and Post-Bachelors) are less likely
to agree (1 in every 4 individual) about the equal educational value of
online learning and classroom teaching, as compared to the lower educated
group (where 1 in every 3 agrees). Similarly, the racial classification of
response shows that the African-Americans are more likely to agree (37\%) as
compared to White (30.9\%) and Other Races (30\%).

The discussion involving the stacked bar graph only presents an association
between the public opinion on the educational value of online learning
relative to in-person classroom teaching
and one covariate at a time, namely, age, previous participation in online
course, education, and race. Such an association can be captured by
regressing the dependent variable on a chosen covariate/regressor. However,
inference based on such an analysis is unlikely to present the true
relationship because there may be other determinants of the dependent
variable which are correlated with the chosen covariate. If ignored, this may
lead to estimation bias and incorrect inferences. For instance, let us
suppose we are interested in the relationship between public opinion on
online learning relative to in-person classroom teaching and the age
category. To this end,  we regress the dependent variable on age category.
However, this relationship is likely to change when we control for previous
participation in online course owing to the correlation between previous
participation in online course and age. To net out such effects and
understand the actual impact of a covariate on the dependent variable, we
next turn to some formal econometric modelling.

\section{Quantile Regression for Binary Outcomes}\label{sec:Model}

Quantile regression, as introduced by \citet{Koenker-Basset-1978}, looks at
quantiles of the (continuous) response variable conditional on the covariates
and thus provides, amongst other things, a comprehensive picture (as compared
to traditional mean regression) of the effect of covariates on the response
variable. Estimation involves minimizing the quantile loss function using
linear programming techniques \citep{KoenkerBook-2005}. Interestingly, the
quantile loss function appears in the exponent of the asymmetric Laplace (AL)
distribution \citep{Yu-Zhang-2005}, which makes minimization of the quantile
loss function equivalent to maximization of the AL likelihood. This
characteristic allowed \citet{Yu-Moyeed-2001} to construct a working
likelihood and propose Bayesian quantile regression. However, when outcomes
are discrete (e.g., binary, ordinal) estimation becomes challenging because
quantiles for discrete outcomes are not readily defined. With discrete
outcomes, the concern is to model the latent utility differential (say,
between making a choice versus not making it or occurrence of an event versus
its non-occurrence) facilitated through the introduction of a latent variable
\citep{Albert-Chib-1993,Greenberg-2012,Rahman-2016}. This applies to both
mean and quantile regressions and is useful for estimation and inference.

Quantile regression for binary outcomes (or binary quantile
regression\footnote{Binary quantile regression is a special of ordinal
quantile regression considered in \citet{Rahman-2016} and can be linked to
the random utility theory in economics \citep{Train-2009,
Jeliazkov-Rahman-2012}. For other developments on Bayesian quantile
regression with discrete outcomes, please see
\citet{Alhamzawi-Ali-Longitudinal2018},
\citet{Alhamzawi-Ali-SingleIndex2018}, \citet{Ghasemzadeh-etal-2018-Comm},
\citet{Ghasemzadeh-etal-2018-METRON}, \citet{Rahman-Vossmeyer-2019},
\citet{Rahman-Karnawat-2019}, \citet{Bresson-etal-2020}.}) was introduced in
\citet{Kordas-2006} and the Bayesian framework was presented in
\citet{Benoit-Poel-2012}. The binary quantile model can be conveniently
expressed using the latent variable $z_{i}$ as follows,
\begin{equation}
\begin{split}
z_{i} & =  x'_{i} \beta_{p}  + \epsilon_{i}, \hspace{0.75in} \forall \;
i=1, \cdots, n, \\
y_{i} & = \left\{ \begin{array}{ll}
1 & \textrm{if} \;  z_{i} > 0,\\
0 & \textrm{otherwise}.
\end{array} \right.
\end{split}
\label{eq:Model1}
\end{equation}
where $x_{i}$ is a $k \times 1$ vector of covariates, $\beta_{p}$ is a $k
\times 1$ vector of unknown parameters at the $p$-th quantile (henceforth,
the subscript $p$ is dropped for notational convenience), $\epsilon_{i}$
follows an AL distribution i.e., $\epsilon_{i} \sim AL(0,1,p)$, and $n$
denotes the number of observations. In our study, the latent variable $z_{i}$
can be interpreted as the latent utility differential between online learning
relative to in-person classroom learning. Whenever the observed response
$y_{i}=1$ (i.e., the respondent answers `Yes' to our question of interest),
propensity to online learning is likely to be high and $z_{i}$ takes a value
in the positive part of the real line. Similarly, when $y_{i}=0$ (i.e., the
respondent answers `No' to our question of interest), the propensity to
online learning is low and $z_{i}$ takes a value in the negative part of the
real line.

We can form a working likelihood from equation~\eqref{eq:Model1} and directly
use it to construct the posterior distribution, but this is not convenient
for MCMC sampling. A preferred alternative is to employ the
normal-exponential mixture of the AL distribution
\citep{Kozumi-Kobayashi-2011}. In this formulation, $\epsilon_{i} = \theta
w_{i} + \tau \sqrt{w_{i}} \,u_{i}$, and the binary quantile model is
re-expressed as,
\begin{equation}
\begin{split}
z_{i} & =  x'_{i} \beta  +  + \theta w_{i} + \tau \sqrt{w_{i}} \,u_{i},
\hspace{0.75in} \forall \; i=1, \cdots, n, \\
y_{i} & = \left\{ \begin{array}{ll}
1 & \textrm{if} \;  z_{i} > 0,\\
0 & \textrm{otherwise}.
\end{array} \right.
\end{split}
\label{eq:Model2}
\end{equation}
where $\theta = \frac{(1-2p)}{p(1-p)}$ and $\tau = \sqrt{\frac{2}{ p(1-p)}}$
are constants, and $w_{i} \sim \mathcal{E}(1)$ is independently distributed
of $u_{i} \sim N(0,1)$. Here, the notations $\mathcal{E}$ and $N$ denote
exponential and normal distributions, respectively. It is clear from
formulation \eqref{eq:Model2} that the latent variable $z_{i}|\beta,w_{i}
\sim N( x'_{i}\beta + \theta w_{i}, \tau^{2} w_{i})$, thus allowing access to
the properties of normal distribution.

\begin{table*}[!b]
\begin{algorithm}[MCMC Algorithm for Binary Quantile Regression]
\label{alg:algorithm1}
\rule{\textwidth}{0.5pt} \small{
\begin{enumerate}
\item    Sample $\beta| z,w$ $\sim$  $N(\tilde{\beta}, \tilde{B})$, where,
\item[]  $\tilde{B}^{-1} = \bigg(\sum_{i=1}^{n}
        \frac{x_{i} x'_{i}}{\tau^{2} w_{i}} + B_{0}^{-1} \bigg) $ \hspace{0.05in}
         and \hspace{0.05in} $\tilde{\beta} = \tilde{B}\bigg( \sum_{i=1}^{n}
         \frac{x_{i}(z_{i} - \theta w_{i})}{\tau^{2} w_{i}} + B_{0}^{-1} \beta_{0}
         \bigg)$.
\item    Sample $w_{i}|\beta, z_{i}$ $\sim$  $GIG \, (0.5, \tilde{\lambda}_{i},
    \tilde{\eta}) $, for $i=1,\cdots,n$, where,
\item[]  $\tilde{\lambda}_{i} = \Big( \frac{ z_{i} - x'_{i}\beta}{\tau}
    \Big)^{2}$ \hspace{0.05in} and \hspace{0.05in} $\tilde{\eta} = \Big(
    \frac{\theta^{2}}{\tau^{2}} + 2 \Big)$.
\item    Sample the latent variable $z|y,\beta,w$ for
         all values of $i=1,\cdots,n$ from an univariate
         truncated normal (TN) distribution as follows,
         \begin{eqnarray*}
         z_{i}|y, \beta,w & \sim & \left\{
         \begin{array}{ll}
         TN_{(-\infty,0]}\bigg(x'_{i}\beta + \theta w_{i},
         \tau^{2} w_{i} \bigg)
         & \textrm{if} \;\;  y_{i} = 0,\\ [0.9em]
         TN_{(0,\infty)}\bigg(x'_{i}\beta + \theta w_{i},
         \tau^{2} w_{i} \bigg)
         & \textrm{if} \;\; y_{i} = 1.
         \end{array}
         \right.
         \end{eqnarray*}
\end{enumerate}}
\rule{\textwidth}{0.5pt}
\end{algorithm}
\end{table*}

By the Bayes' theorem, the complete data likelihood from
equation~\eqref{eq:Model2} is combined with a normal prior distribution on
$\beta$ (i.e., $\beta \sim N(\beta_{0}, B_{0})$) to form the complete data
posterior. This yields the following expression,
\begin{equation}
\begin{split}
\pi(z,\beta,w|y)
          & \propto  \bigg\{ \prod_{i=1}^{n}
          \big[ I(z_{i}>0)I(y_{i}=1) + I(z_{i} \leq 0) I(y_{i}=0) \big]
          \; N(z_{i}|x'_{i}\beta + \theta w_{i}, \tau^{2} w_{i}) \\
          & \times \; \mathcal{E}(w_{i}|1) \bigg\}
          \; N(\beta_{0}, B_{0}).
          \label{eq:CompDataPost}
\end{split}
\end{equation}
The full conditional posterior densities for $(z,\beta, w)$ can be derived
from equation \eqref{eq:CompDataPost} and the model can be estimated using
the Gibbs sampler \citep{Geman-Geman-1984} $-$ a well known MCMC technique
$-$ presented in Algorithm~\ref{alg:algorithm1}. The sampling algorithm is
straightforward and involves sampling $\beta$ conditional on $(z, w)$ from an
updated normal distribution. The latent weight $w$ conditional on $(\beta,z)$
is sampled from a Generalized Inverse Gaussian (GIG) distribution
\citep{Devroye-2014}. Finally, the latent variable $z$ conditional on $(y,
\beta, w)$ is sampled from a truncated normal distribution
\citep{Robert-1995}.

\section{Results} \label{sec:Results}
\begin{sloppypar}
Table \ref{Table:EducValResults} presents the posterior means, and standard
deviations of the parameters from the Bayesian estimation of probit model
\citep{Albert-Chib-1993}, and the binary quantile regression at the 10th,
25th, 50th, 75th and 90th quantiles. We assume the following diffuse prior
distribution: $\beta \sim N(0_{k}, 1000*I_{k})$, where $N$ and $I$ denote a
multivariate normal distribution and an identity matrix of dimension $k$,
respectively. The results are based on 20,000 MCMC iterations after a burn-in
of 5,000 iterations. The inefficiency factors were calculated using the
batch-means method \citep{Greenberg-2012, Chib-2013}. For the five chosen
quantiles, they lie in the range $(6.34, 10.94)$, $(4.18, 5.39)$, $(2.53,
3.16)$, $(2.38, 2.58)$, and $(3.35, 4.24)$. The numbers are small which
indicates a low cost of working with MCMC draws. Trace plots, not shown,
reveal good mixing of the chains. With respect to model comparison measures,
we calculate the conditional log-likelihood, Akaike Information Criterion
(AIC), and the Bayesian Information Criterion (BIC) at the posterior mean.
The corresponding numbers for the probit model is $(-959.43, 1956.86,
2058.93)$, and those for the five quantiles are $(-960.24, 1958.47,
2060.54)$, $(-960.18, 1958.36, 2060.43)$, $(-960.00, 1958.00, 2060.07)$,
$(-959.90, 1957.80, 2059.87)$, and $(-959.16, 1956.33, 2058.40)$.  We also
compute the covariate effects for the statistically significant variables in
the probit model and for each of the five quantiles. These are presented in
Table \ref{Table:EducCE}\footnote{The covariate effects for previous online
course, full-time employment, post-bachelors, bachelors, Northeast and South
are calculated on the respective sub-samples and are a discrete change
compared to their base groups respectively. The covariate effect for female
is calculated on the full sample and is a discrete change compared to male.}
and are calculated marginally of the remaining covariates and the parameters
\citep{Chib-Jeliazkov-2006,Jeliazkov-etal-2008,Jeliazkov-Rahman-2012,Jeliazkov-Vossmeyer-2018,Rahman-Vossmeyer-2019,Bresson-etal-2020}.

While many of the results are in line with extant literature, our results
provide some useful insights into the differences across quantiles. As
previously noted, we are modelling the latent utility differential between
online and in-person classes. The results therefore, may be interpreted as a
utility index of online education. Large positive (negative) values of this
index signify high (low) propensity to favour online classes, and values
around zero would indicate relative indifference between the two
alternatives. A bird's-eye view of the results shows that age, past online
experience, full time employment and gender have a positive effect on the
propensity to favour online education. Higher level of educational degree,
on the other hand, has a negative effect on the willingness towards online
education. We also note some amount of regional variation in the propensity
to favour online classes. To better understand the results, we focus on each
variable separately.
\end{sloppypar}

\begin{table}[!t]
\centering \footnotesize \setlength{\tabcolsep}{2.5pt}
\setlength{\extrarowheight}{2pt}
\setlength\arrayrulewidth{1pt}
\caption{Posterior mean (\textsc{mean}) and standard deviation
(\textsc{std}) of the parameters from the Bayesian estimation of probit
regression
and binary quantile regression.}
\begin{tabular}{p{1.2in} rrr rrr rrr rrr rrrr rr  }
\toprule
& & & & \multicolumn{15}{c}{\textsc{quantile}}   \\
\cmidrule{6-19}
&& \multicolumn{2}{c}{\textsc{probit}} && \multicolumn{2}{c}{\textsc{10th}}
&& \multicolumn{2}{c}{\textsc{25th}} && \multicolumn{2}{c}{\textsc{50th}}
&& \multicolumn{2}{c}{\textsc{75th}} && \multicolumn{2}{c}{\textsc{90th}}   \\
\cmidrule{3-4} \cmidrule{6-7}  \cmidrule{9-10} \cmidrule{12-13}
\cmidrule{15-16}
\cmidrule{18-19}
                   &&  \textsc{mean} & \textsc{std} &&  \textsc{mean} & \textsc{std}
                   &&  \textsc{mean} & \textsc{std} &&  \textsc{mean} & \textsc{std}
                   &&  \textsc{mean} & \textsc{std} &&  \textsc{mean} & \textsc{std}
\\
\midrule
Intercept                && $ -0.80$  & $0.20$  && $-14.02$ & $2.17$  && $ -5.03$
& $0.85$
                         && $ -1.69$  & $0.45$  && $ -0.23$ & $0.46$  && $  1.58$ &
$0.97$  \\
Age/100                  && $  0.58$  & $0.23$  && $  5.39$ & $2.44$  && $  2.24$
& $0.92$
                         && $  1.17$  & $0.49$  && $  1.34$ & $0.52$  && $  3.35$ &
$1.14$  \\
Income/100,000           && $  0.37$  & $0.26$  && $  4.20$ & $2.73$  && $
1.70$ & $1.13$
                         && $  0.92$  & $0.58$  && $  0.87$ & $0.59$  && $  1.02$ &
$1.23$  \\
Sq-Income                && $ -0.28$  & $0.14$  && $ -3.18$ & $1.57$  && $
-1.30$ & $0.65$
                         && $ -0.69$  & $0.33$  && $ -0.63$ & $0.32$  && $ -0.86$ &
$0.64$  \\
Online Course            && $  0.31$  & $0.09$  && $  2.80$ & $0.83$  && $
1.14$ & $0.35$
                         && $  0.64$  & $0.19$  && $  0.75$ & $0.21$  && $  1.74$ &
$0.48$  \\
(Age$<65$)$\ast$ Enroll
                         && $  0.02$  & $0.11$  && $  0.25$ & $1.13$  && $  0.12$ &
$0.43$
                         && $  0.07$  & $0.22$  && $  0.00$ & $0.24$  && $  0.00$ &
$0.51$  \\
Female                   && $  0.14$  & $0.07$  && $  1.72$ & $0.71$  && $  0.71$
& $0.28$
                         && $  0.36$  & $0.15$  && $  0.27$ & $0.15$  && $  0.53$ &
$0.33$  \\
Post-Bachelors           && $ -0.45$  & $0.12$  && $ -4.76$ & $1.32$  && $
-1.95$ & $0.55$
                         && $ -1.02$  & $0.28$  && $ -0.98$ & $0.27$  && $ -2.09$ &
$0.54$  \\
Bachelors                && $ -0.40$  & $0.10$  && $ -4.19$ & $1.21$  && $
-1.69$ & $0.44$
                         && $ -0.90$  & $0.23$  && $ -0.87$ & $0.23$  && $ -1.81$ &
$0.48$  \\
Below Bachelors          && $ -0.09$  & $0.09$  && $ -1.03$ & $0.83$  && $
-0.44$ & $0.35$
                         && $ -0.26$  & $0.19$  && $ -0.17$ & $0.20$  && $ -0.27$ &
$0.47$  \\
Full-time                && $  0.27$  & $0.08$  && $  2.76$ & $0.94$  && $  1.13$
& $0.37$
                         && $  0.58$  & $0.19$  && $  0.58$ & $0.19$  && $  1.28$ &
$0.42$  \\
Part-time                && $  0.17$  & $0.11$  && $  1.67$ & $1.15$  && $  0.68$
& $0.49$
                         && $  0.35$  & $0.24$  && $  0.41$ & $0.24$  && $  0.93$ &
$0.51$  \\
White                    && $ -0.01$  & $0.11$  && $  0.08$ & $1.12$  && $  0.06$
& $0.48$
                         && $  0.03$  & $0.24$  && $ -0.05$ & $0.24$  && $ -0.18$ &
$0.49$  \\
African-American         && $  0.21$  & $0.13$  && $  2.05$ & $1.26$  && $
0.93$ & $0.55$
                         && $  0.48$  & $0.28$  && $  0.43$ & $0.31$  && $  0.87$ &
$0.66$  \\
Urban                    && $ -0.10$  & $0.11$  && $ -1.00$ & $1.12$  && $ -0.36$
& $0.47$
                         && $ -0.23$  & $0.24$  && $ -0.25$ & $0.25$  && $ -0.35$ &
$0.55$  \\
Suburban                 && $  0.06$  & $0.11$  && $  0.58$ & $1.04$  && $
0.32$ & $0.44$
                         && $  0.11$  & $0.23$  && $  0.12$ & $0.24$  && $  0.43$ &
$0.56$  \\
Northeast                && $ -0.29$  & $0.12$  && $ -3.09$ & $1.31$  && $
-1.23$ & $0.56$
                         && $ -0.63$  & $0.28$  && $ -0.63$ & $0.28$  && $ -1.53$ &
$0.59$  \\
West                     && $ -0.06$  & $0.11$  && $ -0.44$ & $1.04$  && $ -0.18$
& $0.42$
                         && $ -0.10$  & $0.22$  && $ -0.18$ & $0.25$  && $ -0.58$ &
$0.56$  \\
South                    && $ -0.22$  & $0.09$  && $ -2.34$ & $0.96$  && $ -0.93$
& $0.39$
                         && $ -0.48$  & $0.20$  && $ -0.46$ & $0.22$  && $ -1.14$ &
$0.51$  \\

\bottomrule
\end{tabular}
\label{Table:EducValResults}
\end{table}

\begin{table}[!t]\centering
\footnotesize \setlength{\tabcolsep}{6pt} \setlength{\extrarowheight}{2pt}
\caption{Covariate Effect.}
\begin{tabular}{lrr rrr rrr }
\toprule
    && &&  \multicolumn{5}{c}{\textsc{quantile}} \\
    \cmidrule{5-9}
&& \textsc{probit}  &&  10th   & 25th    &  50th     &  75th    &  90th     \\
\midrule
Age         && $ 0.0200$
            && $ 0.0175$ & $ 0.0181$ & $ 0.0185$  & $ 0.0202$  &  $ 0.0219$  \\
Online Course
            && $ 0.1088$
            && $ 0.0952$ & $ 0.0971$ & $ 0.1048$  & $ 0.1114$  &  $ 0.1104$  \\
Female      && $ 0.0486$
            && $ 0.0535$ & $ 0.0550$ & $ 0.0549$  & $ 0.0415$  &  $ 0.0351$  \\
Post-Bachelors
            && $-0.1512$
            && $-0.1447$ & $-0.1479$ & $-0.1511$  & $-0.1478$  &  $-0.1395$  \\
Bachelors   && $-0.1356$
            && $-0.1306$ & $-0.1317$ & $-0.1379$  & $-0.1332$  &  $-0.1211$  \\
Full-time   && $ 0.0857$
            && $ 0.0784$ & $ 0.0804$ & $ 0.0813$  & $ 0.0853$  &  $ 0.0874$  \\
Northeast   && $-0.0954$
            && $-0.0920$ & $-0.0914$ & $-0.0918$  & $-0.0940$  &  $-0.1024$  \\
South       && $-0.0777$
            && $-0.0806$ & $-0.0797$ & $-0.0807$  & $-0.0693$  &  $-0.0723$  \\
\bottomrule
\end{tabular}
\label{Table:EducCE}
\end{table}

The coefficient for age is positive across all quantiles. This is not
surprising as online courses invariably attract an older demographic
\citep{Crain-Ragan-2017}. \citet{Goodman-etal-2019} find similar results
highlighting that on average, the online applicants were 34 years of age
compared to 24 years for in–person applicants in their study. Besides, our
result is perhaps indicative of mid-career professionals favouring online
classes since several online courses cater to those active in the workforce,
requiring professional development or retaining by employers
\citep{Kizilcec-etal-2019}. From the calculated covariate effects in Table
\ref{Table:EducCE}, we see that the covariate effect of age is between 1.7 to
2.2 percentage points across the quantiles. Stronger effects are visible in
the upper part of the latent index.

Next, we note that educational value of online classes is favoured positively
by individuals with a full-time employment status compared to base category
(unemployed, students or retired individuals). The coefficient for full-time
employment, compared to the base category, is statistically positive across
the quantiles. The coefficients for part-time employment are
positive but the effects are not statistically different, implying that
regardless of the latent utility for online education, part-time employment
does not impact the decision. Our result for full-time employment is in
agreement with the evidence that demand for online education is high for
employed mid-career professionals, or those who seek professional
development \citep{Simmons-2014,Kizilcec-etal-2019}. It appears to be
commonplace for employers to sponsor their employees’ enrolment into online
courses for training purposes as observed by \citet{Goodman-etal-2019} and
\citet{Deming-etal-2015}.\footnote{The National Post Secondary Student Aid
Study (NPSAS) for 2011-12 that includes a nationally representative
cross-section of institutions and students shows that online students are
older and more likely to be working full-time while enrolled
(\citet{Deming-etal-2015}.} In fact, from Table \ref{Table:EducCE}, the
covariate effect of full-time employment increases the willingness for online
class by 7.8 percentage points in the 10th quantile and consistently
increases across quantiles to about 8.7 percentage points in the 90th
quantile. For individuals who are in the lower part of the latent index,
employment impacts their valuation for online education less than those in
the upper quantile.

Turning to previous exposure to digital learning, we find that individual’s
propensity of valuing online education is higher for those who have had past
participation in online classes for academic credit than those who have not.
We find positive effects of previous exposure to online education across the
quantiles of the utility scale. \citet{Astani-etal-2010,Williams-2006,Goode-2010} show similar
evidence that previous online experience changes the perceptions about an
online learning environment. A positive stance towards online education is
therefore undeniably linked to previous exposure and use of technology. The
covariate effect of previous online class ranges between 9.5 to 11.1
percentage points across the quantiles (See Table \ref{Table:EducCE}).
Although the effect somewhat plateaus at the 75th quantile, our findings
suggest that past online experience increases the probability of valuing
online classes most for those with higher utility for online education.

We also find that females are more in favour of online education relative to
males. This finding is in consonance with \citet{Fortson-etal-2007}, who
propose that female college students are more likely to go online for
communicative and educational purposes while male college students are more
likely to use the internet as a source of entertainment. Perhaps the noted
gender differential could also be a result of differences in past usage of
internet.\footnote{The proportion of individuals with previous experiences of
online education is higher for females in our sample, with 55\% of females
having taken an online course for credit before.} Furthermore, online
education allows for flexible schedules that individuals can customise around
their family and job constraints more easily \citep{Goodman-etal-2019}. This
greater flexibility in schedule likely implies greater willingness for online
education for females. The covariate effect of females, displayed in Table
\ref{Table:EducCE}, shows that being female increases the probability of
valuing online education by 5.3 to 3.5 percentage points from 10th to 90th
quantiles respectively. Strongest effect of female is found in the 25th
quantile and the effect reduces at the 90th quantile. At higher utility,
females are more similar to males than at lower utility.

Next, we find that the coefficient for different levels of education are
consistently negative, relative to the base category (HS and below), across
the quantiles. In each quantile, the post-bachelors category shows a large
negative propensity for online classes vis-à-vis traditional learning. The
effects are also negative for those with a bachelors degree compared to those
with HS education or below. While the effects are negative for
below-bachelors degree, they are not statistically different in comparison to
the base category. This is useful in understanding the differences in
preferences between individuals with different educational qualifications.
Highly educated respondents report diminished value of online classes in
comparison to those with a HS degree or below. This points to some
degree of stigma towards online education as the level of educational
qualification rises \citep{Kizilcec-etal-2019}. The decrease in utility from
online classes is likely driven by greater intensity of learning and teaching
at graduate or post graduate levels. Students perceive better learning from
face-to-face interactions, and visualising materials. The self-regulatory
nature of physical classroom teaching perhaps enables students to track their
understanding of the course. Our result finds support in
\citet{Chen-etal-2013} who note more favourable student outcomes for
traditional classrooms versus an online mode for advanced accounting courses,
and \citet{Otter-etal-2013} who note that students believe that they must do
the teaching and learning on their own in online courses in contrast to what
they feel about the time and effort from faculty for traditional courses.
This perhaps reduces the value they attach to online education at higher
degree levels. According to \citet{ONeill-Sai-2014}, traditional classes also
allow for better relationship and lines of communication with the instructor.
Other studies have shown that students face difficulty in keeping up
motivation in online classes. This is likely to become more prominent at
higher levels of education. Findings documented in
\citet{Anstine-Skidmore-2005} and \citet{Fendler-etal-2011}, suggest that
students perform worse in online courses compared to face-to-face classes at
undergraduate and graduate levels also support our results of lower
willingness for online education by them.\footnote{Individual’s prior
experience with online courses and their performance likely play a role in
determining the value they attach to online education. As per
\citet{Parker-etal-2011}, roughly 39\% of those who have taken an online
course before respond favourably to online educational value whereas about
27\% of those with no prior online education favourably value online
classes.}  Examining our covariate effects from Table \ref{Table:EducCE}, we
see that the effects are negative and vary between 13.9 to 15.1 percentage
points for individuals with a post-bachelors degree. The result from the 50th
quantile is similar to the probit result implying a 15.1 percentage points
reduction in the probability of valuing online education. The covariate
effects are slightly lower for those with a bachelors degree and range
between 12.1 to 13.7 percentage point decline in their opinion about online
classes.

Looking at geographical locations, we note some significant regional effects
in driving the opinion about online education. We find negative effects for
those living in the Northeast, the South, and the West relative to the
Midwest (the omitted category). However, the effects are not statistically
important for those residing in the West. The largest negative effects are
seen in the Northeast across all quantiles, followed by the South in
comparison to the Midwest.\footnote{Use of technology in the Mid-west and the
South is higher compared to the East. In fact, the East coasters are found to
lag behind the rest of the country in some aspects of technology adoption as
per \citet{Parker-etal-2011}.}. Regional differences in the popularity of
online courses are likely driven by use of technology, student population,
course design and support provided by the universities, as well as the
philosophies of universities in the region. Educational value of online
courses are thus considered higher in regions where online education is more
popular.\footnote{\citet{Miller-2014} states that universities in Arizona are
considered to be early adopters of online teaching techniques, in fact
preferring faculty with experience in technology.} \citet{Xu-Jaggars-2013}
highlight the importance of the institutional state in determining the
cultural capital around technology. \cite{Allen-Seaman-2007} also suggest
that the Southern states represented over one-third of total online
enrolments in 2005-06 and the proportion of Southern institutions with fully
online programs is steadily rising.\footnote{Regions with students having
high levels of technological proficiency are more likely to take courses
which integrate technology, major in technology-rich disciplines, and pursue
technology-rich careers \citep{Xu-Jaggars-2013}.} Our covariate effect
calculations indicate that in the Northeast, the probability of favouring
online classes reduces by 9.1 to 10.2 percentage points relative to the
Midwest, across the quantiles. The highest negative effect is found for those
in the 90th quantile. The propensity to value online education is reduces by
6.9 to 8 percentage points in the Southern regions compared to the Midwest.

We also examine the effect of race and find no noteworthy racial differences
in the willingness towards online education relative to in-person education.
Specifically, the coefficients for White, relative to the base category
(Other Races), are statistically not different from zero. Similarly, the
coefficients for African-Americans across the quantiles are statistically
equivalent to zero. This likely indicates that, after controlling for
different educational levels and previous exposure to online learning,
individuals across racial groups seem to hold similar attitudes about the
online classes as an educational tool. Our results fall in line with
\citep{Cotten-Jelenewicz-2006,Odell-etal-2000,Jones-etal-2009,Bowen-etal-2014},
who note that the digital divide upheld by race may be narrowing, and in some
cases negligible among college campuses in the US. Contrary to this,
\citet{Figlio-etal-2013}, find negative outcomes for Hispanic students.

The coefficient for urban areas are negative and for the suburban areas are
positive compared to the base group of rural residents. However, the effects
are not statistically important implying that area of residence does not
seem to impact the opinion about online classes. Interestingly, income levels
have a positive effect across the quantiles but once again the effects are
not statistically different from zero. Although, \citet{Horrigan-Rainie-2006}
find that affluent families have better access to internet, we find no
convincing evidence that income plays a role in determining opinion about
online classes vis-à-vis traditional classes.

\section{Conclusion} \label{sec:Conclusion}
Technological advancements and the rising cost of higher education have
rendered online education as an attractive substitute or a complementary
technique for teaching and learning. With the online enrolment growth rate in
the US at 9.3 percent, over 6.7 million students were estimated to have taken
at least one online course in 2012 \citep{Allen-Seaman-2013}. Considering
this trend, in this paper, we examine \textit{public opinion} about the value
of online learning methods in comparison to in-person education across United
States. Evaluating public opinion can not only serve as a guide for policy,
but also help design the transformative shift away from physical classrooms
as the dominant paradigm of teaching and learning. Some scholars argue that
public policies should in fact be guided by public opinion, so that mass
opinion and democracy is upheld \citep{Monroe-1998,Paletz-2013}.

Extant literature points to uncertainty regarding the quality and rigour of
online education. While there is some degree of adaptation to specific online
courses offered by traditional universities as blended learning, reservations
about fully online degree programs remain.
As such, we assert that the approach provided in this paper leads to a richer
view of how the demographic covariates may influence public opinion about the
educational value of online classes, thereby, better informing future
educational policies. We find important effects of employment status,
previous online experience, age and gender on the propensity towards online
education across the latent utility scale. We also note that willingness
towards online classes versus in-person classes is lower for highly educated
individuals. While interesting regional variations exist, we find no evidence
of race and income on the propensity for online education.

\begin{sloppypar}
We conclude with three main questions for future work. First, creating an
in-depth, systematic support for both faculty and students, in transitioning
from traditional to online teaching-learning platforms, is not an inexpensive
venture. With considerable fixed costs incurred in training, course creations
and delivery methods for online education
\citep{Ginn-Hammond-2012,Xu-Jaggars-2013}, what would be the incentives to
switch back to in-person classrooms or blended formats in a post-pandemic
general equilibrium? Second, evidence suggests that online education
democratises and improves access to education \citep{Goodman-etal-2019}.
Given this, it may be interesting to examine the trade-off between a
perceived decrease in outcomes and efficacy of online education versus the
increase in the exposure of education to previously inaccessible population
from a policy perspective. Finally, it is well established that upward
mobility in teaching colleges are largely influenced by student feedback and
evaluations \citep{Chen-Hoshower-2003,Mcclain-2018,Krautmann-1999}. If we
consider the current health landscape across the globe as a period of
deviation from the true nature of dynamics between education and technology,
one ought to think about how student feedback during this phase of
aberration, will contribute to upward mobility of faculty.
\end{sloppypar}

\clearpage \pagebreak
\pdfbookmark[1]{References}{unnumbered}       

\bibliography{EducValuebib}

\begin{thebibliography}{}
\newcommand{\enquote}[1]{``#1''}

\bibitem[Albert and Chib(1993)Albert and Chib]{Albert-Chib-1993}
Albert, J.~H. and Chib, S. (1993), \enquote{Bayesian Analysis of Binary and
  Polychotomous Response Data,} \emph{Journal of the American Statistical
  Association}, 88, 669--679.

\bibitem[Alhamzawi and Ali(2018a)Alhamzawi and
  Ali]{Alhamzawi-Ali-Longitudinal2018}
Alhamzawi, R. and Ali, H. T.~M. (2018a), \enquote{Bayesian Quantile Regression
  for Ordinal Longitudinal Data,} \emph{Journal of Applied Statistics}, 45,
  815--828.

\bibitem[Alhamzawi and Ali(2018b)Alhamzawi and
  Ali]{Alhamzawi-Ali-SingleIndex2018}
Alhamzawi, R. and Ali, H. T.~M. (2018b), \enquote{Bayesian Single-Index
  Quantile Regression for Ordinal Data,} \emph{Communications in Statistics $-$
  Simulation and Computation}, pp. 1--15.

\bibitem[Allen and Seaman(2007)Allen and Seaman]{Allen-Seaman-2007}
Allen, I.~E. and Seaman, J. (2007), \enquote{Making the Grade: Online Education
  in the {U}nited {S}tates, 2006,} \textit{Sloan Consortium}, Feb 2007.
  Available: https://files.eric.ed.gov/fulltext/ED530101.pdf [Last accessed: 11
  July 2020].

\bibitem[Allen and Seaman(2011)Allen and Seaman]{Allen-Seaman-2011}
Allen, I.~E. and Seaman, J. (2011), \enquote{Going the Distance: Online
  Education in the {U}nited {S}tates, 2011.} \textit{Sloan Consortium}, Nov
  2011. Available: https://files.eric.ed.gov/fulltext/ED529948.pdf [Last
  accessed: 11 July 2020].

\bibitem[Allen and Seaman(2013)Allen and Seaman]{Allen-Seaman-2013}
Allen, I.~E. and Seaman, J. (2013), \enquote{Changing Course: Ten Years of
  Tracking Online Education in the {U}nited {S}tates,} \textit{Sloan
  Consortium}, Jan 2013. Available:
  https://files.eric.ed.gov/fulltext/ED541571.pdf [Last accessed: 11 July
  2020].

\bibitem[Allen and Seaman(2016)Allen and Seaman]{Allen-Seaman-2016}
Allen, I.~E. and Seaman, J. (2016), \enquote{Online Report Card: Tracking
  Online Education in the {U}nited {S}tates,} \textit{Babson Survey Research
  Group}, Feb 2016. Available: https://files.eric.ed.gov/fulltext/ED572777.pdf
  [Last accessed: 11 July 2020].

\bibitem[Alpert et~al.(2016)Alpert, Couch, and Harmon]{Alpert-etal-2016}
Alpert, W.~T., Couch, K.~A., and Harmon, O.~R. (2016), \enquote{A Randomized
  Assessment of Online Learning,} \emph{American Economic Review}, 106,
  378--82.

\bibitem[Anstine and Skidmore(2005)Anstine and Skidmore]{Anstine-Skidmore-2005}
Anstine, J. and Skidmore, M. (2005), \enquote{A Small Sample Study of
  Traditional and Online Courses with Sample Selection Adjustment,} \emph{The
  Journal of Economic Education}, 36, 107--127.

\bibitem[Astani et~al.(2010)Astani, Ready, and Duplaga]{Astani-etal-2010}
Astani, M., Ready, K.~J., and Duplaga, E.~A. (2010), \enquote{Online Course
  Experience Matters: Investigating Students’ Perceptions of Online
  Learning,} \emph{Issues in Information Systems}, 11, 14--21.

\bibitem[Benoit and Poel(2012)Benoit and Poel]{Benoit-Poel-2012}
Benoit, D.~F. and Poel, D. V.~D. (2012), \enquote{Binary Quantile Regression: A
  {B}ayesian Approach based on the Asymmetric {L}aplace Distribution,}
  \emph{Journal of Applied Econometrics}, 27, 1174--1188.

\bibitem[Bettinger et~al.(2017)Bettinger, Fox, Loeb, and
  Taylor]{Bettinger-etal-2017}
Bettinger, E.~P., Fox, L., Loeb, S., and Taylor, E.~S. (2017), \enquote{Virtual
  Classrooms: How Online College Courses Affect Student Success,}
  \emph{American Economic Review}, 107, 2855--2875.

\bibitem[Bowen et~al.(2014)Bowen, Chingos, Lack, and Nygren]{Bowen-etal-2014}
Bowen, W.~G., Chingos, M.~M., Lack, K.~A., and Nygren, T.~I. (2014),
  \enquote{Interactive Learning Online at Public Universities: Evidence from a
  Six-campus Randomized Trial,} \emph{Journal of Policy Analysis and
  Management}, 33, 94--111.

\bibitem[Bresson et~al.(2020)Bresson, Lacroix, and Rahman]{Bresson-etal-2020}
Bresson, G., Lacroix, G., and Rahman, M.~A. (2020), \enquote{Bayesian Panel
  Quantile Regression for Binary Outcomes with Correlated Random Effects: An
  Application on Crime Recidivism in Canada,} \emph{Empirical Economics},
  (forthcoming).

\bibitem[Cassens(2010)Cassens]{Cassens-2010}
Cassens, T.~S. (2010), \enquote{Comparing the Effectiveness of Online and
  Face-to-Face Classes among California Community College Students,} Ph.D.
  thesis, University of Southern California.

\bibitem[Chen et~al.(2013)Chen, Jones, and Moreland]{Chen-etal-2013}
Chen, C.~C., Jones, K.~T., and Moreland, K.~A. (2013), \enquote{Online
  Accounting Education versus In-class Delivery: Does Course Level Matter?}
  \emph{Issues in Accounting Education}, 28, 1--16.

\bibitem[Chen and Fu(2009)Chen and Fu]{Chen-Fu-2009}
Chen, S.-Y. and Fu, Y.-C. (2009), \enquote{Internet Use and Academic
  Achievement: Gender Differences in Early Adolescence,} \emph{Adolescence},
  44, 797--812.

\bibitem[Chen and Hoshower(2003)Chen and Hoshower]{Chen-Hoshower-2003}
Chen, Y. and Hoshower, L.~B. (2003), \enquote{Student Evaluation of Teaching
  Effectiveness: An Assessment of Student Perception and Motivation,}
  \emph{Assessment \& Evaluation in Higher Education}, 28, 71--88.

\bibitem[Chib(2013)Chib]{Chib-2013}
Chib, S. (2013), \enquote{Introduction to Simulation and {MCMC} Methods,} in
  \emph{The {O}xford Handbook of {B}ayesian Econometrics}, eds. J.~Geweke,
  G.~Koop, and H.~V. Dijk, pp. 183--218, Oxford University Press, Oxford.

\bibitem[Chib and Jeliazkov(2006)Chib and Jeliazkov]{Chib-Jeliazkov-2006}
Chib, S. and Jeliazkov, I. (2006), \enquote{Inference in Semiparametric Dynamic
  Models for Binary Longitudinal Data,} \emph{{Journal of the American
  Statistical Association}}, 101, 685--700.

\bibitem[Cooper(2006)Cooper]{cooper2006digital}
Cooper, J. (2006), \enquote{The Digital Divide: The Special case of Gender,}
  \emph{Journal of Computer Assisted Learning}, 22, 320--334.

\bibitem[Cotten and Jelenewicz(2006)Cotten and
  Jelenewicz]{Cotten-Jelenewicz-2006}
Cotten, S.~R. and Jelenewicz, S.~M. (2006), \enquote{A Disappearing Digital
  Divide among College Students? Peeling away the Layers of the Digital
  Divide,} \emph{Social Science Computer Review}, 24, 497--506.

\bibitem[Crain and Ragan(2017)Crain and Ragan]{Crain-Ragan-2017}
Crain, S.~J. and Ragan, K.~P. (2017), \enquote{Online Versus Face-to-Face
  Course Learning Effectiveness: Measured Outcomes for Intermediate Financial
  Management,} \emph{Journal of Financial Education}, 43, 243--261.

\bibitem[Deming et~al.(2015)Deming, Goldin, Katz, and
  Yuchtman]{Deming-etal-2015}
Deming, D.~J., Goldin, C., Katz, L.~F., and Yuchtman, N. (2015), \enquote{Can
  Online Learning bend the Higher Education Cost Curve?} \emph{American
  Economic Review}, 105, 496--501.

\bibitem[Devroye(2014)Devroye]{Devroye-2014}
Devroye, L. (2014), \enquote{Random Variate Generation for the Generalized
  Inverse Gaussian Distribution,} \emph{Statistics and Computing}, 24,
  239--246.

\bibitem[Fendler et~al.(2011)Fendler, Ruff, and Shrikhande]{Fendler-etal-2011}
Fendler, R.~J., Ruff, C., and Shrikhande, M. (2011), \enquote{Online versus
  In-class Teaching: Learning Levels Explain Student Performance,}
  \emph{Journal of Financial Education}, 37, 45--63.

\bibitem[Figlio et~al.(2013)Figlio, Rush, and Yin]{Figlio-etal-2013}
Figlio, D., Rush, M., and Yin, L. (2013), \enquote{Is it Live or is it
  Internet? Experimental Estimates of the Effects of Online Instruction on
  Student Learning,} \emph{Journal of Labor Economics}, 31, 763--784.

\bibitem[Fortson et~al.(2007)Fortson, Scotti, Chen, Malone, and
  Del~Ben]{Fortson-etal-2007}
Fortson, B.~L., Scotti, J.~R., Chen, Y.-C., Malone, J., and Del~Ben, K.~S.
  (2007), \enquote{Internet Use, Abuse, and Dependence Among Students at a
  Southeastern Regional University,} \emph{Journal of American College Health},
  56, 137--144.

\bibitem[Geman and Geman(1984)Geman and Geman]{Geman-Geman-1984}
Geman, S. and Geman, D. (1984), \enquote{Stochastic Relaxation, Gibbs
  Distributions, and the Bayesian Restoration of Images,} \emph{IEEE
  Transactions on Pattern Analysis and Machine Intelligence}, 6, 721--741.

\bibitem[Ghasemzadeh et~al.(2018a)Ghasemzadeh, Ganjali, and
  Baghfalaki]{Ghasemzadeh-etal-2018-METRON}
Ghasemzadeh, S., Ganjali, M., and Baghfalaki, T. (2018a), \enquote{Bayesian
  Quantile Regression for Analyzing Ordinal Longitudinal Responses in the
  Presence of Non-ignorable Missingness,} \emph{METRON}, 76, 321--348.

\bibitem[Ghasemzadeh et~al.(2018b)Ghasemzadeh, Ganjali, and
  Baghfalaki]{Ghasemzadeh-etal-2018-Comm}
Ghasemzadeh, S., Ganjali, M., and Baghfalaki, T. (2018b), \enquote{Bayesian
  Quantile Regression for Joint Modeling of Longitudinal Mixed Ordinal
  Continuous Data,} \emph{Communications in Statistics $-$ Simulation and
  Computation}, pp. 1--21.

\bibitem[Ginn and Hammond(2012)Ginn and Hammond]{Ginn-Hammond-2012}
Ginn, M.~H. and Hammond, A. (2012), \enquote{Online Education in Public
  Affairs: Current State and Emerging Issues,} \emph{Online Journal of Distance
  Learning Administration}, 18, 247--270.

\bibitem[Goode(2010)Goode]{Goode-2010}
Goode, J. (2010), \enquote{Mind the Gap: The Digital Dimension of College
  Access,} \emph{The Journal of Higher Education}, 81, 583--618.

\bibitem[Goodman et~al.(2019)Goodman, Melkers, and Pallais]{Goodman-etal-2019}
Goodman, J., Melkers, J., and Pallais, A. (2019), \enquote{Can Online Delivery
  Increase Access to Education?} \emph{Journal of Labor Economics}, 37, 1--34.

\bibitem[Greenberg(2012)Greenberg]{Greenberg-2012}
Greenberg, E. (2012), \emph{Introduction to {Bayesian} Econometrics}, 2nd
  Edition, Cambridge University Press, New York.

\bibitem[Horrigan and Rainie(2006)Horrigan and Rainie]{Horrigan-Rainie-2006}
Horrigan, J.~B. and Rainie, L. (2006), \enquote{The Internet's growing Role in
  Life's Major Moments,} \textit{Pew Internet \& American Life Project, Pew
  Research Center}, 19 April 2006. Available:
  https://www.pewinternet.org/wp-content/uploads/sites/9/media/Files/Reports/2006/PIP{\_}Major-Moments{\_}2006.pdf.pdf
  [Last accessed: 11 July 2020].

\bibitem[Jeliazkov and Rahman(2012)Jeliazkov and Rahman]{Jeliazkov-Rahman-2012}
Jeliazkov, I. and Rahman, M.~A. (2012), \enquote{Binary and Ordinal Data
  Analysis in Economics: Modeling and Estimation,} in \emph{Mathematical
  Modeling with Multidisciplinary Applications}, ed. X.~S. Yang, pp. 123--150,
  John Wiley \& Sons Inc., New Jersey.

\bibitem[Jeliazkov and Vossmeyer(2018)Jeliazkov and
  Vossmeyer]{Jeliazkov-Vossmeyer-2018}
Jeliazkov, I. and Vossmeyer, A. (2018), \enquote{The Impact of Estimation
  Uncertainty on Covariate Effects in Nonlinear Models,} \emph{Statistical
  Papers}, 59, 1031--1042.

\bibitem[Jeliazkov et~al.(2008)Jeliazkov, Graves, and
  Kutzbach]{Jeliazkov-etal-2008}
Jeliazkov, I., Graves, J., and Kutzbach, M. (2008), \enquote{Fitting and
  Comparison of Models for Multivariate Ordinal Outcomes,} \emph{Advances in
  Econometrics: {Bayesian} Econometrics}, 23, 115--156.

\bibitem[Jones et~al.(2009)Jones, Johnson-Yale, Millermaier, and
  P{\'e}rez]{Jones-etal-2009}
Jones, S., Johnson-Yale, C., Millermaier, S., and P{\'e}rez, F.~S. (2009),
  \enquote{US College Students’ Internet Use: Race, Gender and Digital
  Divides,} \emph{Journal of Computer-Mediated Communication}, 14, 244--264.

\bibitem[Joyce et~al.(2015)Joyce, Crockett, Jaeger, Altindag, and
  O'Connell]{Joyce-etal-2015}
Joyce, T., Crockett, S., Jaeger, D.~A., Altindag, O., and O'Connell, S.~D.
  (2015), \enquote{Does Classroom Time Matter?} \emph{Economics of Education
  Review}, 46, 64--77.

\bibitem[Kirtman(2009)Kirtman]{Kirtman-2009}
Kirtman, L. (2009), \enquote{Online versus In-class Courses: An Examination of
  Differences in Learning Outcomes,} \emph{Issues in Teacher Education}, 18,
  103--116.

\bibitem[Kizilcec et~al.(2019)Kizilcec, Davis, and Wang]{Kizilcec-etal-2019}
Kizilcec, R., Davis, D., and Wang, E. (2019), \enquote{Online Degree Stigma and
  Stereotypes: A new Instrument and Implications for Diversity in Higher
  Education,} \emph{Available at SSRN 3339768}.

\bibitem[Koenker(2005)Koenker]{KoenkerBook-2005}
Koenker, R. (2005), \emph{Quantile Regression}, Cambridge University Press,
  Cambridge.

\bibitem[Koenker and Bassett(1978)Koenker and Bassett]{Koenker-Basset-1978}
Koenker, R. and Bassett, G. (1978), \enquote{Regression Quantiles,}
  \emph{Econometrica}, 46, 33--50.

\bibitem[Kordas(2006)Kordas]{Kordas-2006}
Kordas, G. (2006), \enquote{Smoothed Binary Regression Quantiles,}
  \emph{Journal of Applied Econometrics}, 21, 387--407.

\bibitem[Kozumi and Kobayashi(2011)Kozumi and Kobayashi]{Kozumi-Kobayashi-2011}
Kozumi, H. and Kobayashi, G. (2011), \enquote{Gibbs Sampling Methods for
  {Bayesian} Quantile Regression,} \emph{Journal of Statistical Computation and
  Simulation}, 81, 1565--1578.

\bibitem[Krautmann and Sander(1999)Krautmann and Sander]{Krautmann-1999}
Krautmann, A.~C. and Sander, W. (1999), \enquote{Grades and Student Evaluations
  of Teachers,} \emph{Economics of Education Review}, 18, 59--63.

\bibitem[Krieg and Henson(2016)Krieg and Henson]{Krieg-Henson-2016}
Krieg, J.~M. and Henson, S.~E. (2016), \enquote{The Educational Impact of
  Online Learning: How do University Students perform in Subsequent Courses?}
  \emph{Education Finance and Policy}, 11, 426--448.

\bibitem[McClain et~al.(2018)McClain, Gulbis, and Hays]{Mcclain-2018}
McClain, L., Gulbis, A., and Hays, D. (2018), \enquote{Honesty on Student
  Evaluations of Teaching: Effectiveness, Purpose, and Timing Matter!}
  \emph{Assessment \& Evaluation in Higher Education}, 43, 369--385.

\bibitem[Miller(2014)Miller]{Miller-2014}
Miller, M.~D. (2014), \emph{Minds online}, Harvard University Press, Cambridge.

\bibitem[Monroe(1998)Monroe]{Monroe-1998}
Monroe, A.~D. (1998), \enquote{Public Opinion and Public Policy, 1980-1993,}
  \emph{Public Opinion Quarterly}, 62, 6--28.

\bibitem[Norum and Weagley(2006)Norum and Weagley]{Norum-Weagley-2006}
Norum, P.~S. and Weagley, R.~O. (2006), \enquote{College Students, Internet
  Use, and Protection from Online Identity Theft,} \emph{Journal of Educational
  Technology Systems}, 35, 45--63.

\bibitem[Odell et~al.(2000)Odell, Korgen, Schumacher, and
  Delucchi]{Odell-etal-2000}
Odell, P.~M., Korgen, K.~O., Schumacher, P., and Delucchi, M. (2000),
  \enquote{Internet Use among Female and Male College Students,}
  \emph{CyberPsychology \& Behavior}, 3, 855--862.

\bibitem[O’Neill and Sai(2014)O’Neill and Sai]{ONeill-Sai-2014}
O’Neill, D.~K. and Sai, T.~H. (2014), \enquote{Why not? Examining College
  Students’ Reasons for Avoiding an Online Course,} \emph{Higher Education},
  68, 1--14.

\bibitem[Otter et~al.(2013)Otter, Seipel, Graeff, Alexander, Boraiko, Gray,
  Petersen, and Sadler]{Otter-etal-2013}
Otter, R.~R., Seipel, S., Graeff, T., Alexander, B., Boraiko, C., Gray, J.,
  Petersen, K., and Sadler, K. (2013), \enquote{Comparing Student and Faculty
  Perceptions of Online and Traditional Courses,} \emph{The Internet and Higher
  Education}, 19, 27--35.

\bibitem[Paletz et~al.(2013)Paletz, Owen, and Cook]{Paletz-2013}
Paletz, D.~L., Owen, D.~M., and Cook, T.~E. (2013), \emph{American Government
  and Politics in the Information Age}, Flat World Knowledge.

\bibitem[Parker et~al.(2011)Parker, Lenhart, and Moore]{Parker-etal-2011}
Parker, K., Lenhart, A., and Moore, K. (2011), \enquote{The Digital Revolution
  and Higher Education: College Presidents, Public Differ on Value of Online
  Learning.} \textit{Pew Internet \& American Life Project, Pew Research
  Center}, 28 Aug 2011. Available:
  https://files.eric.ed.gov/fulltext/ED524306.pdf [Last accessed: 11 July
  2020].

\bibitem[Rahman(2016)Rahman]{Rahman-2016}
Rahman, M.~A. (2016), \enquote{Bayesian Quantile Regression for Ordinal
  Models,} \emph{Bayesian Analysis}, 11, 1--24.

\bibitem[Rahman and Karnawat(2019)Rahman and Karnawat]{Rahman-Karnawat-2019}
Rahman, M.~A. and Karnawat, S. (2019), \enquote{Flexible Bayesian Quantile
  Regression in Ordinal Models,} \emph{Advances in Econometrics}, 40B,
  211--251.

\bibitem[Rahman and Vossmeyer(2019)Rahman and Vossmeyer]{Rahman-Vossmeyer-2019}
Rahman, M.~A. and Vossmeyer, A. (2019), \enquote{Estimation and Applications of
  Quantile Regression for Binary Longitudinal Data,} \emph{Advances in
  Econometrics}, 40B, 157--191.

\bibitem[Robert(1995)Robert]{Robert-1995}
Robert, C.~P. (1995), \enquote{Simulation of Truncated Normal Variables,}
  \emph{Statistics and Computing}, 5, 121--125.

\bibitem[Simmons(2014)Simmons]{Simmons-2014}
Simmons, G.~R. (2014), \enquote{Business statistics: A Comparison of Student
  Performance in Three Learning Modes,} \emph{Journal of Education for
  Business}, 89, 186--195.

\bibitem[Train(2009)Train]{Train-2009}
Train, K. (2009), \emph{Discrete Choice Methods with Simulation}, Cambridge
  University Press, Cambridge.

\bibitem[Williams(2006)Williams]{Williams-2006}
Williams, S.~L. (2006), \enquote{The Effectiveness of Distance Education in
  Allied Health Science Programs: A Meta-analysis of Outcomes,} \emph{The
  American Journal of Distance Education}, 20, 127--141.

\bibitem[Xu and Jaggars(2013)Xu and Jaggars]{Xu-Jaggars-2013}
Xu, D. and Jaggars, S.~S. (2013), \enquote{The Impact of Online Learning on
  Students’ Course Outcomes: Evidence from a Large Community and Technical
  College System,} \emph{Economics of Education Review}, 37, 46--57.

\bibitem[Yu and Moyeed(2001)Yu and Moyeed]{Yu-Moyeed-2001}
Yu, K. and Moyeed, R.~A. (2001), \enquote{Bayesian Quantile Regression,}
  \emph{Statistics and Probability Letters}, 54, 437--447.

\bibitem[Yu and Zhang(2005)Yu and Zhang]{Yu-Zhang-2005}
Yu, K. and Zhang, J. (2005), \enquote{A Three Parameter Asymmetric {Laplace}
  Distribution and its Extensions,} \emph{Communications in Statistics --
  Theory and Methods}, 34, 1867--1879.

\end{thebibliography}
\bibliographystyle{jasa}



\end{document}